# A Lightweight Forum-based Distributed Requirement Elicitation Process for Open Source Community

[1,5]Han Lai, *[2]Rong Peng, [3]Dong Sun, [4]Jia Liu
[1,*2,3,4] *State Key Lab. of Software Engineering, Wuhan University, Wuhan, China*
[5]*College of Computer Science, Chongqing Technology and Business University, Chongqing, China*
[1,5]laihan_ctbu@126.com, [2]rongpeng@sklse.org, [3]sundong@whu.edu.cn,
[4]liujia8111@yahoo.com.cn

**Abstract**
*Nowadays, lots of open source communities adopt forum to acquire scattered stakeholders' requirements. But the requirements collection process always suffers from the unformatted description and unfocused discussions. In this paper, we establish a framework ReqForum to define the meta-model of the requirement elicitation forum. Based on it, we propose a lightweight forum-based requirements elicitation process which includes six steps: template-based requirements creation, opinions collection, requirements collection, requirements management, capability identification and the incentive mechanism. According to the proposed process, the prototype SKLSEForum is established by composing the Discuz and its existed pulg-ins. The implementation indicates that the process is feasible and the cost is economic.*

## 1. Introduction

At present, a lot of open source communities adopt forum to collect the distributed stakeholders' opinions or suggestions [1], such as Unix [2], Mozilla [3] and Android [4], etc. However, these forums only focus on providing platforms for opinion exchange. The existence of high randomness of thread creation and large number of redundant threads threatens the utilization of the valuable information in the platform [5].

In this paper, we firstly establish a framework ReqForum to define the meta-model of the requirements elicitation forum, which can be used by open source communities to collect scattered stakeholders' requirements. Based on the framework, we propose a requirement elicitation process which can effectively support the distributed requirements elicitation.

The rest of this paper is organized as follows: Section 2 presents related work. Section 3 presents the forum-based framework. Section 4 proposes the forum enhanced requirements elicitation process. Section 5 describes the implementation of the prototype. Section 6 concludes with an outlook on future work.

## 2. Related work

Open Source Software Development (OSSD) represents a collaborative community based effort to develop software in which the users participate in deciding what features to build, and a subgroup of developers participate in designing the solution, writing code, and deploying and maintaining the system[1]. Initial studies of requirements development across multiple types of OSSD projects [1,21] find that OSS product requirements are continuously emerging and asserted after they have been implemented, rather than relatively stable and elicited before being implemented [22,23].

Many enterprises and research institutes try to develop the lightweight CASE tools to support distributed requirements elicitation and analysis [24-27]. The process proposed in this paper is inspired by the work of Cleland-Huang, et al. [5-9], who explore and evaluate the forum-based requirements gathering and prioritization processes by vendor-based open source software projects. The emphasis of their work focuses on how to utilize data mining and machine learning techniques to automatically group stakeholders' ideas into forums, and use recommendation technologies to help promote these forums to potentially stakeholders. The main focus of our work is on how to promote the effectiveness of the stakeholders' participation and stimulate their enthusiasm. The process proposed includes the following six aspects: reducing information redundancy, clarifying the thread's content and processing status, promoting the negotiation among stakeholders, stimulate the desire of participation and identifying the capability of the stakeholders.

## 3. Forum-based requirement elicitation framework

The framework defines the meta-model of the requirements elicitation forum which facilitates the asynchronous communication among stakeholders (See Figure 1).

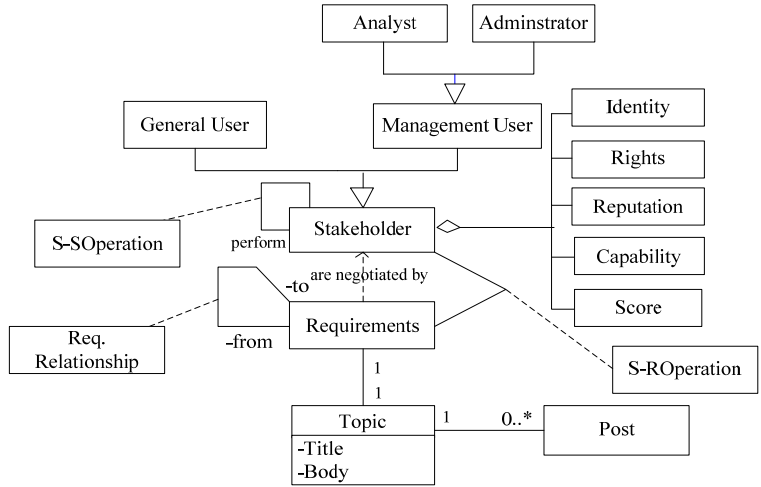

**Figure 1.** Forum-based requirements elicitation framework ReqForum

In ReqForum, stakeholders are classified into General Users and Management Users. The stakeholders using the open source community's forum to give opinions or suggestions are named as General Users. Requirements analysts and the forum system administrators are regarded as Management Users. Each type of stakeholders has their own Identity, Rights, Reputation, Capability and Score in the forum. Stakeholders with different types have different Rights to use forum's functions; the stakeholder's Reputation represents the trusted degree and also represents its prestige in the forum; the stakeholder's Capability represents the capability of identifying valuable requirements, which is related with the familiarity with the system function or the mastery of domain knowledge; the stakeholder's Score represents the active degree of participating in the forum's discussion. The Requirements are represented as threads. A requirement consist of a Topic and multiple Posts. A general Topic consists of Title and Body. The Req. Relation represents the relationship, which exists between the requirements. The S-ROperation represents the stakeholders can perform various operations on the forum's Topic and Post to reflect the activity of different stakeholders involved in requirements elicitation. The S-SOperation represents the operations between Stakeholders, e.g. Management User rewards score to general users in the forum to encourage discussion, General Stakeholder communicates with each other by sending and receiving message, etc.

## 4. Enhanced forum-based requirement elicitation process

To support the forum-based opinion and suggestion's collection, we propose a requirement elicitation process based on the Reqforum framework. The features of requirements elicitation process lie in the following six aspects: requirements creation, opinions collection, requirements collection, requirements management, stakeholder's capability identification and the incentive mechanism of forum (See Figure 2).

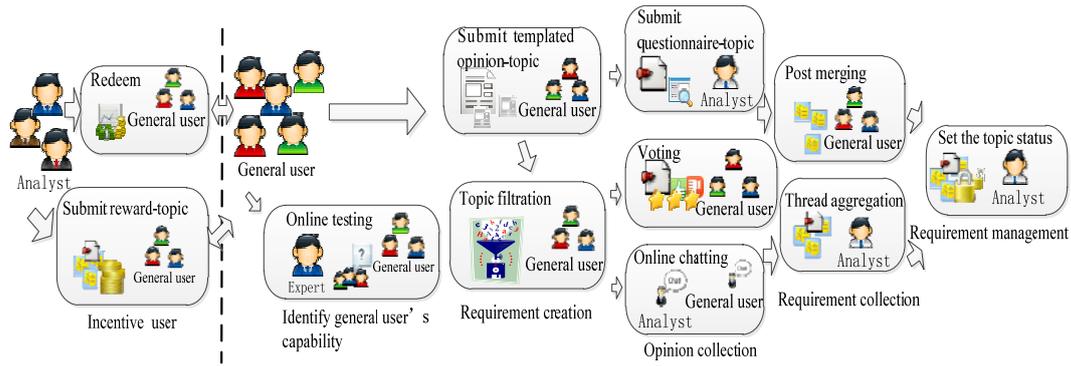

**Figure 2**. An overview of requirements elicitation process based on the enhanced forum

### 4.1. Requirements creation

In traditional forum system, general user submits a requirement description in a topic. But the topic always has no format constraints, which leads to the poor quality of requirements description. In addition, general users often intentionally or unintentionally repeat the same topic to attract the attention of Analysts. It inevitably results in duplication of the discussion, and makes the redundant content in the forum system [5]. Therefore, we introduce template mechanism to standardize the opinion topic's creation, and use the unsupervised spam detection method based on string alienness measures [10] to filter the repeated topics (See Figure 3).

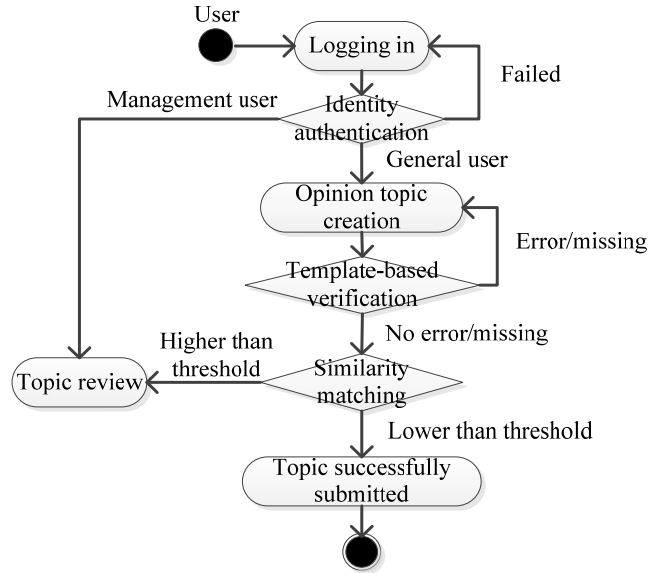

**Figure 3.** Activity diagram for requirements creation

The forum user must login first; after identity authentication, the user can create a new opinion topic according to the guidance of requirements description template; and then, the topic must pass the template-based verification; when he want to submit it, a comparison between its content and all previous topics will be carried out. If the similarity degree is lower than a specific threshold, the topic can be successfully submitted.

The template-based verification can help stakeholders describe their requirements clearly and completely. We extend the Topic to Opinion Topic, Questionnaire Topic (See section 4.2), and Reward Topic (see section 4.6), as illustrated in Figure 4. Opinion Topic is based on the requirement template (Req. Template). Each kind of templates consists of a series of Template Items. Item Relationship can be established between Template Items. Item can be divided into Mandatory Item and Optional Item according to its property.

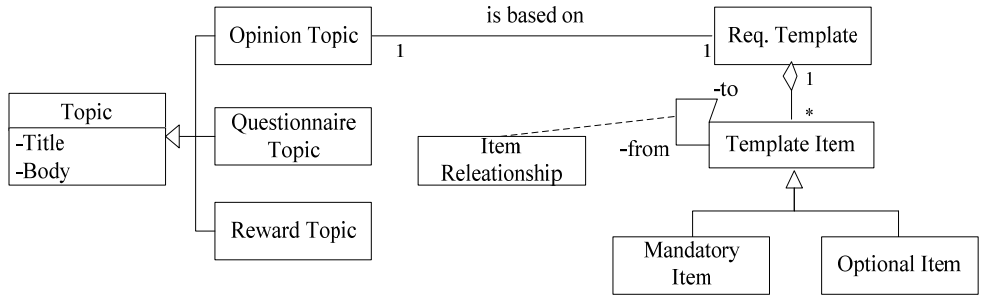

**Figure 4.** Template structure of Topic

### 4.2. Opinions collection

In traditional forum system, stakeholders' posts always suffer from high redundancy and unfocused discussion. In addition, the posts are simple and vague [11]. Therefore, we collect the users' opinions according to the following manners:
- Introducing questionnaire topic to collect the stakeholders' opinions, namely, Analysts can collect General Users' feedback in the form of questionnaire.
- Introducing voting mechanism to collect the priority and preference opinions;
- Introducing the online chatting to facilitate Analysts to organize online negotiation among opinion authors and related stakeholders.

### 4.3. Requirements collection

Requirements creation and opinions collection focus on collecting the original users' opinions and suggestions. Requirements collection focus on provide facilities to help analysts refine users' requirements from the topics and posts. The shortage of using traditional forum to elicit stakeholders' requirements is that the users' requirements assessment opinions are separated in the posts and cannot be sorted easily. A combination mechanism of posts is proposed to settle the problem. Namely, if the current post and the previous post are submitted by the same user, the posts will be merged automatically to reduce the information redundancy and facilitate the analysis (See Figure 5). Besides, we utilize the thread aggregation function to semi-automatically help the analysts deal with the opinion topic and its posts in the form of single list page, to avoid numerous posts spread in different pages (See Figure 6).

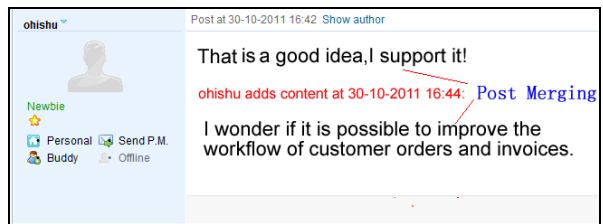

**Figure 5.** The screenshot of posts merging

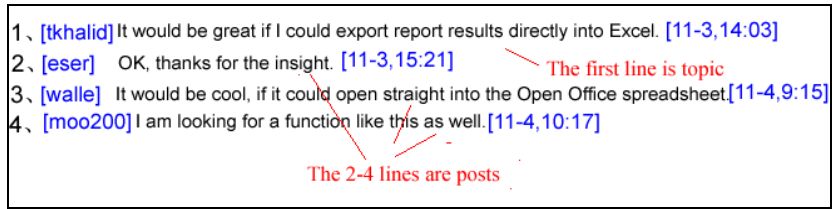

**Figure 6.** The screenshot of threads aggregation

### 4.4. Requirements management

The traditional forum lacks of effective progress tracking mechanism, and the perception of requirements status is weak [5]. We introduce the opinion-topic management process to settle the problem (See Figure 7).

The opinion topic has the following states: New, Suggestion collected, Negotiation, Unlocked, Locked and Cancelled. The transition flow among these states is as follows:

- When stakeholder submits an opinion topic, its status is "New" (1);
- After the topic being created, it will be open to all stakeholders to collect comments and suggestions, its status change to "Suggestion collected" (2);
- If great differences exist among the posts of the topic, analysts can convene related stakeholders to negotiate through integrated online chatting or questionnaire investigation. Then the topic enters the "Negotiation" status (3). On the other hand, if the stakeholders' opinions are consistent, the requirement enters "Locked" status directly (8).
- If the "Negotiation" reaches consistent, the topic will be in "Locked" status (4); otherwise, it may be changed into "Cancelled" status to indicate that the requirement will be not realized; (5) or reenter the "Suggestion collected" status to collect more information (9) ;
- When the topic is "Locked", analysts have the right to "Unlocked" if necessary (6), and organize related stakeholders to "Negotiation" again (7).
- If the "New" topic representing certain requirement repeats other topics or gets low evaluation in "Suggestion collected" , the analysts have the right to decide whether to "Cancelled" it or not (10, 11).

The states' marks will be displayed with the topics to make the stakeholders clearly know the status of requirements processing.

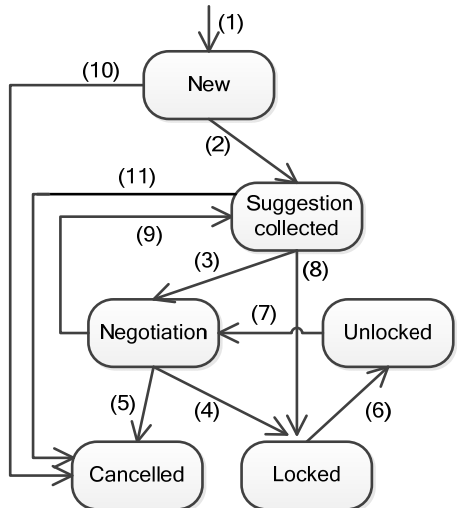

**Figure 7.** The state transition diagram of requirements states change

### 4.5. Stakeholders' capability identification

Finding domain experts from the general users is very important in distributed requirements elicitation [12]. It can improve the quality of requirements negotiation and requirements analysis. We introduce the capability assessment mechanism. By integrating the online test page, the management user can set online test. If the new registered general user logs in the forum, his capability can be identified by performing specified test, and then, the corresponding rights will be assigned to him according to the test result.

### 4.6. Encouragement to stakeholders

To encourage general users to participate in the discussion, the encouragement mechanism is necessary. We add the reward topic and redeem mechanism. The management users can submit the reward topic to reward Q & A, and general users can answer the questions to prompt their reputation and increase scores, and then redeem the gift by their scores.

## 5. Prototype implementation

We have implemented the SKLSEForum based on the Discuz (www.discuz.net), a popular free/open source forum solution, and its plug-ins to illustrate that the proposed process is feasible and the implementation cost is economic. Table 1 lists Discuz and its plug-ins which are used to support the specific aspects in the process.

**Table 1.** The process supported by Discuz and its plugins

| Forum-based distributed requirement elicitation process | Enhanced forum-based feature | Discuz | Plugins |
|---|---|---|---|
| Requirements creation | Template opinion-topic submission | √ | |
| | Topic filtration | | Anti-Spam Plugin[13] |
| Opinions collection | Questionnaire-topic submission | | Questionnaire Plugin[14] |
| | Voting | | Voting Plugin[15] |
| | Online chatting | | Chatting Plugin[16] |
| Requirements collection | Post merging | | Posts Merging Plugin[17] |
| | Thread aggregation | | Thread Aggregation Plugin[18] |
| Requirements management | Topic states setting | √ | |
| Stakeholders' capability identification | Online testing | | Online testing Plugin[19] |
| Encouragement to stakeholders | Redeeming | | Redeem Gift Plugin[20] |
| | Reward-topic submission | √ | |
| "√" represents the feature supported by Discuz | | | |

## 6. Conclusion and future work

This paper presents a requirements elicitation process based on forum to enhance forum's capability of eliciting requirements. The process includes six steps with which it can provide the ability of reducing information redundancy, clarifying the thread's content and processing status, promoting the negotiation among stakeholders, encouraging the desire of participation and identifying the capability of the stakeholders.

In future, more efforts will be paid on conducting case studies in both academic and industrial environments to assess and improve the applicability and usability of our prototype.

## 7. Acknowledgements

The research was supported by the National Natural Science Foundation of China under Grant No. 61170026, 60940028 and 60703009, the Outstanding Youth Foundation of Hubei Province under Grant No. 2009CDA148, the Youth Chenguang Science Project of Wuhan (200950431189) and the National Basic Research Program of China (973) under Grant No. 2007CB310801.